\documentclass[useAMS,usenatbib]{mn2e}
\usepackage{epsfig,amsmath,amssymb,amsfonts,mathrsfs,latexsym,graphicx}
\include{journals}

\usepackage{times}

\newcommand{\Msun}{\ensuremath{\textrm{M}_{\odot}}}
\newcommand{\Rsun}{\ensuremath{\textrm{R}_{\odot}}}

\newcommand{\kms}{km\hspace{0.25em}s$^{-1}$}

\newcommand{\SiI}{\mbox{Si\hspace{0.25em}{\sc i}}}
\newcommand{\SiII}{\mbox{Si\hspace{0.25em}{\sc ii}}}

\newcommand{\CaII}{\mbox{Ca\hspace{0.25em}{\sc ii}}}

\newcommand{\FeII}{\mbox{Fe\hspace{0.25em}{\sc ii}}}
\newcommand{\FeIII}{\mbox{Fe\hspace{0.25em}{\sc iii}}}

\newcommand{\NiII}{\mbox{Ni\hspace{0.25em}{\sc ii}}}
\newcommand{\Fefs}{$^{56}$Fe}
\newcommand{\Feff}{$^{54}$Fe}
\newcommand{\Cofs}{$^{56}$Co}
\newcommand{\Nifs}{$^{56}$Ni}
\newcommand{\Nife}{$^{58}$Ni}

\newcommand{\MCh}{$M_{\textrm{Ch}}$}
\newcommand{\KE}{\ensuremath{E_{\rm K}}}
\newcommand{\Dm}{\ensuremath{\Delta m_{15}(B)}}

\newcommand{\aap}{A\&A}

\newcommand{\apj}{ApJ}
\newcommand{\apjs}{ApJS}
\newcommand{\apjl}{ApJ}
\newcommand{\aj}{AJ}

\newcommand{\mnras}{MNRAS}
\newcommand{\na}{New Astronomy}
\newcommand{\nat}{Nature}

\newcommand{\pasa}{PASA}

\newcommand{\sci}{Sci}

\newcommand{\eg}{e.g.,\ }

\def\gsim{\mathrel{\rlap{\lower 4pt \hbox{\hskip 1pt $\sim$}}\raise 1pt \hbox {$>$}}}
\def\lsim{\mathrel{\rlap{\lower 4pt \hbox{\hskip 1pt $\sim$}}\raise 1pt \hbox {$<$}}}

\begin{document}

\title[SN\,2011fe: nebular spectra]
  {Nebular spectra and abundance tomography of the type Ia supernova SN\,2011fe:
  a normal SN Ia with a stable Fe core}
\author[P. A.~Mazzali et al.]{P. A. Mazzali$^{1,2}$
\thanks{E-mail: P.Mazzali@ljmu.ac.uk}, 
M. Sullivan$^{3}$, 
A. V. Filippenko$^{4}$, 
P. M. Garnavich$^{5}$, 
\newauthor K. I. Clubb$^{4}$, 
K. Maguire$^{6}$, 
Y.-C. Pan$^{7}$, 
B. Shappee$^{8}$,
J. M. Silverman$^{9}$, 
S. Benetti$^{10}$, 
\newauthor   
S. Hachinger$^{11,12}$, K. Nomoto$^{13}$, and E. Pian$^{14,15}$
\\
\\
  $^{1}$Astrophysics Research Institute, Liverpool John Moores University, IC2, Liverpool Science Park, 146 Brownlow Hill, Liverpool L3 5RF, UK\\
  $^{2}$Max-Planck-Institut f\"ur Astrophysik, Karl-Schwarzschild-Str. 1, D-85748 Garching, Germany\\
  $^{3}$School of Physics and Astronomy, University of Southampton, Southampton, SO17 1BJ, UK\\
  $^{4}$Department of Astronomy, University of California, Berkeley, CA 94720-3411, USA\\
  $^{5}$Department of Physics, University of Notre Dame, Notre Dame, IN 46556, USA\\
  $^{6}$ESO, Karl-Schwarzschild-Str. 2, D-85748 Garching, Germany\\
  $^{7}$Department of Astronomy, University of Illinois, Urbana, IL 61801, USA \\
  $^{8}$Carnegie Observatories, 813 Santa Barbara Street, Pasadena, California, 91101 USA \\
  $^{9}$Department of Astronomy, University of Texas, Austin, TX 78712-1205, USA \\
  $^{10}$Istituto Nazionale di Astrofisica-OAPd, vicolo dell'Osservatorio 5, 35122 Padova, Italy\\
  $^{11}$Institut f\"ur Mathematik, Universit\"at W\"urzburg, Emil-Fischer-Str. 30, 97074 W\"urzburg, Germany\\ 
  $^{12}$Institut f\"ur Theoretische Physik und Astrophysik, Universit\"at W\"urzburg, Emil-Fischer-Str. 31, 97074 W\"urzburg, Germany\\
  $^{13}$Kavli IPMU, University of Tokyo, Kashiwano-ha 5-1-5, Kashiwa, Chiba 277-8583, Japan\\
  $^{14}$Institute of Space Astrophysics and Cosmic Physics, via P. Gobetti 101, 40129 Bologna, Italy \\
  $^{15}$Scuola Normale Superiore, Piazza dei Cavalieri 7, I-56126 Pisa, Italy \\
}

\date{Accepted ... Received ...; in original form ...}
\pubyear{2014}
\volume{}
\pagerange{}

\maketitle

\begin{abstract} 
A series of optical and one near-infrared nebular spectra covering the first
year of the Type Ia supernova SN\,2011fe are presented and modelled. The 
density profile that proved best for the early optical/ultraviolet spectra, 
``$\rho$-11fe'', was extended to lower velocities to include the regions that 
emit at nebular epochs. 
Model $\rho$-11fe is intermediate between the fast deflagration model W7 and a 
low-energy delayed-detonation. Good fits to the nebular spectra are obtained if 
the innermost ejecta are dominated by neutron-rich, stable Fe-group species, 
which contribute to cooling but not to heating. 
The correct thermal balance can thus be reached for the strongest [\FeII] and 
[\FeIII] lines to be reproduced with the observed ratio. 
The \Nifs\ mass thus obtained is $0.47 \pm 0.05\,\Msun$. The bulk of 
\Nifs\ has an outermost velocity of $\sim 8500$\,\kms. The mass of stable iron 
is $0.23 \pm 0.03\, \Msun$. Stable Ni has low abundance, 
$\sim 10^{-2}\, \Msun$. This is sufficient to reproduce an observed emission 
line near 7400\,\AA. A sub-Chandrasekhar explosion model with mass  
$1.02\, \Msun$ and no central stable Fe does not reproduce the observed line 
ratios. A mock model where neutron-rich Fe-group species are located above 
\Nifs\ following recent suggestions is also shown to yield spectra that are 
less compatible with the observations.
The densities and abundances in the inner layers obtained from the nebular 
analysis, combined with those of the outer layers previously obtained, are used 
to compute a synthetic bolometric light curve, which compares favourably with 
the light curve of SN\,2011fe. 
\end{abstract}


\section{Introduction}
\label{sec:introduction}

Given the importance of Type Ia supernovae (SNe~Ia), one fundamental question
concerns their origin. There is strong evidence that they are the explosion of 
compact stars, most likely carbon-oxygen (CO) white dwarfs (WDs). Such stars
can undergo thermonuclear explosive burning which results in the production of
radioactive \Nifs, whose decay powers the SN luminous output, and other
elements along the $\alpha$-chain. Apart from \Nifs, the element which is
produced most abundantly is silicon, whose presence characterises the spectra
of SNe\,Ia at early times \citep{nugent11,folatelli12,parrent12}. We know that
\Nifs\ is produced in variable amounts \citep{mazzali07a}, and that the
relation between \Nifs, luminosity, and opacity \citep{pintoeast00,mazzali01}
shapes the light curves of SNe\,Ia, which are to first order a one-parameter
family where broader light curves belong to more luminous SNe
\citep{phillips93}. The explosion is thought to start when the WD reaches a
central temperature of $\sim 10^9$\,K, which happens when its mass approaches
the Chandrasekhar limit, or it may be triggered by accretion or merger events
in lower-mass WDs. 

What we do not know for sure, despite much effort, is how the CO-WD reaches
such a large mass, and how it then explodes. We need to establish this so that 
we can then tackle the problem of how the explosion can produce SNe with
different \Nifs\ content and hence different luminosity.  Popular models for
pushing the WD mass close to the Chandrasekhar limit involve transfer of mass
from a nondegenerate binary companion (main sequence or red giant), the
single-degenerate (SD) scenario, or the merger of two WDs, the
double-degenerate (DD) scenario. Both scenarios have pros and cons. 

The SD scenario ensures repeatability, a constant exploding mass, and a spread
of ages, but it is supposed to involve hydrogen accretion, which should lead to
the detection of H, either in the outer layers as a leftover of the accreting
material or at low velocities as is expected of H stripped from the companion
by the SN ejecta \citep{marietta00,pakmor11}.  However, only rarely has H been
directly detected in SNe\,Ia \citep{Silverman13}. Interestingly, in these cases
much H is seen, such as in SN\,2002ic  \citep{hamuy2003,deng2004} or PTF11kx
\citep{dilday12}. Also, the identification of a surviving companion is not
certain \citep{ruiz-lapuente04,kerzendorf09}. 

The DD scenario may occur more frequently. It is, however, unclear how many
merger events actually exceed a combined mass of about the Chandrasekhar mass.
The proposed scnearios, violent mergers \citep{pakmor11} or collisions
\citep{rosswog09,thompson11,kushnir13}, produce diverse outcomes depending on 
the initial conditions. The question then is whether they look like a typical 
SNe\,Ia, especially in view of the fact that they involve large asymmetries
because ignition happens at the point of contact rather than at the highest
density. 

In a further twist, the possibility that WDs below the Chandrasekhar limit also
explode if they accrete He-rich matter from a companion has also been proposed
\citep{livnearnett95}. This, the sub-Chandra scenario, may again lead to a
different set of observed properties \citep[\eg][]{nugent97,woosleykasen11}. 

Since all of these scenarios are based on solid physical ideas, it is perfectly
possible that they are all actually verified in nature. Is this indeed the
case? Does this add to the uncertainty in our use of SNe\,Ia as distance
indicators? While the latter is a question that can also be addressed by using
large samples of SNe, the former is one that requires the accurate study of a
few, well-observed SNe. 

The biggest differences among explosion models are expected to be found either
at high velocities (presence of circumstellar matter, amount of material at
high velocity as an indication of kinetic energy, metal content of the
progenitor, unburned material from the progenitor or the companion), or at the
lowest ones (central densities, composition, material dragged from companion
upon impact). Different epochs in a supernova's evolution yield different
information. Early-time data can be used to infer  the size of the progenitor,
possibly that of the companion (in the SD scenario), and the properties of the
outer layers of the exploding star, which may not be affected by thermonuclear
burning and may thus carry information about the accreting material.
Observations over the first month or so tell us about general properties: the
light-curve shape (mass, opacity), ejecta velocity (kinetic energy), and
composition of the outer layers (extent of \Nifs,  intermediate-mass elements
(IMEs)). Correlations have been found among these quantities, appearing to
suggest some regularity. 

However, if information about the central regions is required, one has to wait
for the nebular phase. At late times ($\sim 1$\,yr), opacities are low and it
is possible to explore the lowest-velocity material. This corresponds to the
highest densities, and hence it should bear the trace of the ignition and the
outcome of the explosion.  

SN\,2011fe \citep{nugent11} is the perfect candidate for accurate studies. When
it was discovered it was the nearest SN\,Ia in almost 40\,yr. It has been
observed extensively at frequencies from the radio \citep{horesh12,chomiuk12}, 
to the far-infrared (IR) with {\it Herschel} \citep{johansson13}, to the mid-IR
with {\it Spitzer} \citep{mcclelland13}, to the near-IR \citep{matheson12}, to
the optical
\citep{nugent11,li11,richmond12,munari13,shappee13,pereira13,tsvetkov13}, to
the ultraviolet (UV) with {\it Swift} \citep{nugent11,brown12} and with the HST
\citep{foleykirsh13,mazzali14}, to the X-rays with {\it Swift}
\citep{horesh12,margutti12} and {\it Chandra} \citep{margutti12}, and finally
to the gamma rays \citep{isern13}.

SN\,2011fe was a normal SN\,Ia hosted in a spiral galaxy, M101. The early
discovery of the SN by the PTF collaboration allowed a good estimate to be made
of the time of explosion (\citealt{nugent11}, revised by \citealt{pironakar13}
and \citealt{mazzali14}), and also made it possible to estimate the size  of
the exploding star to be $R_\textrm{prog}\lesssim 0.02$\,\Rsun\ 
\citep{bloom12}.  The early-time light curve did not reveal the signature
expected from a shocked companion, contrary to predictions if this were a
sizable star \citep{nugent11,kasen10}. Early-time spectra exhibit traces of
carbon in the outer layers \citep{parrent12,mazzali14} and are consistent with
the low metallicity of the host galaxy \citep{mazzali14}. The spectra have been
claimed to be consistent with both a SD and a DD scenario \citep{roepke12}. In
the outer layers, a density intermediate between the fast deflagration model W7
and a low-energy delayed detonation was found to be the best solution for the
optical/UV spectra \citep{mazzali14}. Merger models and sub-Chandrasekhar-mass
models seem to have lower densities in the inner layers, and a nebular study
can help determining how well they might reproduce the properties of SNe\,Ia.

In the following, the nebular-phase data for SN\,2011fe are presented (Section
2). We show models of the optical spectra in Section 3 using the density
profile derived from the early-time spectra. In Section 4, we extend the models
to the near-infrared (NIR) and discuss alternative models. A discussion
(Section 5) and conclusion (Section 6) follow.

\section{Nebular spectra of SN\,2011fe}

A series of 10 nebular spectra of SN\,2011fe was taken over the period March to
August 2012. An observing log can be found in Table~\ref{tab:obslog}.

Four spectra were taken with the William Herschel Telescope (WHT) using the
Intermediate dispersion Spectrograph and Imaging System (ISIS). ISIS is a
dual-arm spectrograph; we used the R158R grating and GG495 order blocker (red
arm), as well as the R300B grating (blue arm) and the 5300\,\AA\  dichroic,
giving a wavelength coverage of $\sim 3500$--10,000\,\AA. Four spectra were
taken using the Lick Observatory 3-m Shane telescope and the Kast dual-channel
spectrograph \citep{miller93}. The 600/4310 grism was used in the blue arm and
the 300/7500 grating in the red arm, giving a wavelength coverage of
3450--10,200\,\AA.

All of the Lick spectra were reduced using standard techniques 
\citep[\eg][]{Silverman12}.  Routine CCD processing and spectrum extraction
were completed with IRAF\footnote{IRAF is distributed by the National Optical
Astronomy Observatories, which are operated by the Association of Universities
for Research in Astronomy, Inc., under cooperative agreement with the National
Science Foundation (NSF).}. We obtained the wavelength scale from low-order
polynomial fits to calibration-lamp spectra. Small wavelength shifts were then
applied to the data after cross-correlating a template sky to the night-sky
lines that were extracted along with the SN. Using our own IDL routines, we fit
spectrophotometric standard-star spectra to the data in order to flux calibrate
our spectra and to remove telluric lines \citep{Wade88, Matheson00}.

Spectra of SN\,2011fe were obtained with the Large Binocular Telescope (LBT)
and Multi-Object Dual Spectrograph \citep[MODS;][]{pogge12} on three nights in
2012: March 24, April 27, and June 12 (UT dates are used throughout this
paper).  Details of the data acquisition and reduction can be found in
\citet{shappee13}.

NIR spectra were also obtained with the LBT LUCIFER spectrograph
\citep{seifert03} on 2012 May 1, close to the date of one of the MODS optical
spectra. We used the 210 line\,mm$^{-1}$ grating and $1.0''$-wide slit,
resulting in a resolution of 4000 in the $J$ band. Data were taken at three
grating tilts to obtain spectra in the $J$, $H$, and $K$ bands. Consecutive
exposures were dithered along the slit and used to subtract the sky background.
Images at each slit position were then combined and extracted using the IRAF
``twodspec'' package, and the two spectra were averaged to create the final
spectrum. The spectra were wavelength calibrated using night-sky emission
lines, and telluric absorption features were removed using a spectrum of the
A0~V star HIP67848 taken immediately after the SN spectra. The NIR magnitudes
of HIP67848 were then used to flux calibrate the spectra of SN\,2011fe.

An acquisition image of SN\,2011fe was obtained in the $J$ band with LUCIFER.
Using two 2MASS catalog stars in the field we estimate the brightness of the SN
to be $J=17.20\pm 0.06$\,mag.

\begin{table}
\caption{Observing log for the nebular spectra of SN\,2011fe}
\label{tab:obslog}
\centering
\begin{tabular}{llll}
UT Date & Telescope & Instrument & Wavelength\\
&&&Coverage\\
\hline
2012 Mar 2  & WHT  & ISIS & 3500--10,000\,\AA\ \\
2012 Mar 24 & LBT  & MODS & 3300--10,000\,\AA\ \\
2012 Apr 2  & Lick & Kast & 3450--10,300\,\AA\ \\
2012 Apr 23 & Lick & Kast & 3450--10,200\,\AA\ \\
2012 Apr 27 & LBT  & MODS & 3300--10,000\,\AA\ \\
2012 May 1  & LBT  & Lucifer & 1.17--1.31\,$\micron$ \\
2012 May 1  & LBT  & Lucifer & 1.55--1.74\,$\micron$ \\
2012 May 1  & LBT  & Lucifer & 2.05--2.37\,$\micron$ \\
2012 May 26 & WHT  & ISIS & 3500--9500\,\AA\ \\
2012 Jun 12 & LBT  & MODS & 3300--10,000\,\AA\ \\
2012 Jun 25 & WHT  & ISIS & 3500--10,600\,\AA\ \\
2012 Jul 17 & Lick & Kast & 3450--10,300\,\AA\ \\
2012 Aug 21 & WHT  & ISIS & 3500--10,000\,\AA\ \\
2012 Aug 23 & Lick & Kast & 3450--10,200\,\AA\ \\
\hline
\end{tabular}
\end{table}

The spectra were reduced using custom-written pipelines based on standard
procedures in IRAF and IDL, broadly following the reduction procedures outlined
by \citet{ellis08}, including flux calibration and telluric-feature removal. 
All spectra were observed at the parallactic angle \citep{filippenko82} in
photometric conditions.  ``Error'' spectra are derived from a knowledge of the
CCD properties and Poisson statistics, and are tracked throughout the reduction
procedure. The spectra are also rebinned (in a weighted way) to a common
dispersion, and the two arms of the spectrographs joined.

The nebular spectra of SN \,2011fe have been checked against the late-time 
broad-band photometry of \citet{tsvetkov13}. The broad-band magnitudes at the
epochs of the nebular spectra were derived, for each band, from a linear fit of
the available photometric points on the exponentially declining tail of the
light curve. The flux calibration of the spectra was then checked against the
photometry using the IRAF task {\sc stsdas.hst\_calib.synphot.calphot}.  Next,
we locked the absolute flux scale of each spectrum, normalising the synthesised
$BVRI$ magnitudes from the spectrum to the interpolated broad-band magnitudes.
The 2012 Aug. 21 and 2012 Aug. 23 spectra were combined into a single spectrum
to improve the signal-to-noise ratio.

\section{Modelling}

\begin{table*}
\caption{Nebular models}
\label{tab:models}
\centering
\begin{tabular}{ccccccccc}
Date & Epoch & $M$(CO) & $M$(IME) & $M$(\Nifs) & $M$(\Feff) & $M$(\Nife) & $M$(NSE) & $v$(1-zone) \\
 & [rf days] & $\Msun$ & $\Msun$ & $\Msun$ & $\Msun$ & $\Msun$ & $\Msun$ & km\,s$^{-1}$ \\
\hline
2012 Mar 2 & 192 & 0.24 & 0.42 & 0.41 & 0.29 & 0.007 & 0.71 & 8000 \\
2012 Apr 2 & 223 & 0.24 & 0.42 & 0.44 & 0.26 & 0.007 & 0.71 & 7900 \\
2012 Apr 23& 244 & 0.24 & 0.42 & 0.45 & 0.25 & 0.006 & 0.70 & 7900 \\
2012 Apr 27& 248 & 0.24 & 0.42 & 0.46 & 0.24 & 0.004 & 0.70 & 7800 \\
2012 May 1 & 252 & 0.24 & 0.41 & 0.48 & 0.22 & 0.004 & 0.69 & 7800 \\
2012 May 26& 277 & 0.24 & 0.41 & 0.46 & 0.22 & 0.007 & 0.70 & 7900 \\
2012 Jun 25& 307 & 0.24 & 0.41 & 0.48 & 0.21 & 0.007 & 0.70 & 7600 \\
2012 Jul 17& 329 & 0.24 & 0.39 & 0.51 & 0.22 & 0.007 & 0.73 & 7500 \\
2012 Aug 22& 364 & 0.24 & 0.38 & 0.54 & 0.18 & 0.008 & 0.72 & 7700 \\
\hline
\end{tabular}
\end{table*}

We modelled the nebular spectra of SN\,2011fe using our well-tested SN nebular
code. The code is based on approximations developed by \citet{axelrod80}, and
it has been developed and discussed in various papers
\citep{mazzali01,mazzali07b}. The SN ejecta are divided into a number of
spherical shells, whose density reflects that of an explosion model and whose
composition can be varied arbitrarily. The ejecta size and density are rescaled
to the epoch of the desired model assuming homologous expansion. \Nifs\ decay
provides the heating. In the first step of the procedure, $\gamma$-rays and
positrons are emitted according to the distribution of \Nifs. Their propagation
is followed with a Monte Carlo scheme \citep{cappellaro97}. While $\gamma$-rays
can travel a long way before being absorbed in the late-time low-density SN
ejecta,  positrons mostly deposit their energy locally. Deposition of the
energy carried by $\gamma$-rays and positrons determines the heating rate.
Under nebular conditions, this collisional heating is balanced by cooling via
line emission. The ionisation and excitation of relevant species is computed
assuming non-local thermodynamic equilibrium (NLTE), balancing heating and
cooling in an iterative scheme. When convergence is reached, line emissivity is
transformed into an emission-line spectrum. An emitted spectrum is computed for
every shell, and the final emerging spectrum is the sum of the contributions
from all shells. 

\begin{figure}
\includegraphics[width=83mm]{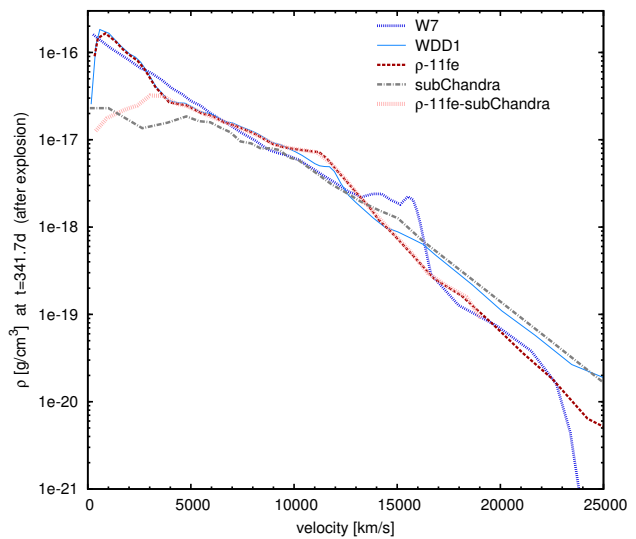}
\caption{Density structures used in this paper: W7 (blue), delayed-detonation
model WDD1 (light blue, solid), sub-Chandra model CDT3 (grey), $\rho$-11fe (red),
 and a sub-Chandra $\rho$-11fe
model with $M = 1.2\, \Msun$ (light red).}
\label{fig:densities}
\end{figure}

A successful model should be able to reproduce the evolution of the spectrum
with time, reflecting the decreasing heating rate and deposition. 
\citet{mazzali14} showed that the early-time spectra of SN\,2011fe cannot be
satisfactorily reproduced with available SN\,Ia explosion models, and proposed
a density distribution characterised by a small but nonzero mass at high
velocities, above $\sim$\,23,000\,\kms\  (see Fig. \ref{fig:densities}).  This
model was shown to be able to reproduce both the optical and UV spectra
simultaneously. The early-time study could not explore the deepest parts of the
ejecta, which are optically thick at early epochs. Here we extend the results
of \citet{mazzali14} and explore the deepest regions of the ejecta.

\citet{mazzali14} concluded that in SN\,2011fe some $\sim 0.4\, \Msun$ of
\Nifs\ are located at $v \gsim 4500$\,\kms. Estimates of the total \Nifs\ mass
for SN\,2011fe suggest a value close to $0.5\, \Msun$ \citep{pereira13}. If
correct, this implies that only a small fraction of all \Nifs\ is located at
low velocities. If the total ejected mass of SN\,2011fe was close or equal to
the Chandrasekhar mass, using the density structure $\rho$-11fe leaves $\sim
0.25\, \Msun$ below 4500\,\kms. Additionally, because of limitations of our
early-time code caused by the lower boundary black-body assumption
\citep[\eg][]{mazzali93}, abundances in layers where \Nifs\ is dominant may not
be fully reliable. In SN\,2011fe this affects the region below $\sim
7000$\,\kms. In any case, we adopted the abundances determined by
\citet{mazzali14} as our initial distribution, and adapted those in the region
below 4500\,\kms\ in order to fit the spectra. The density structure of model
$\rho$-11fe, as well as those of other models used in this paper, are shown in
Figure \ref{fig:densities}.

Assuming that the spectra are properly calibrated in flux (there may be some
uncertainty when spectra were taken at different telescopes, see below), the
overall flux level should reflect the total mass of \Nifs. An important remark
here is that even in the case of SN\,2011fe the uncertainty in flux calibration
can account for an uncertainty in the luminosity level of up to $\sim 10$ per
cent. In the modelling, we adopted for SN\,2011fe a distance of 6.4\,Mpc
(distance modulus $\mu = 29.04$\,mag; \citealt[][find $\mu = 29.04 ± 0.05 $ 
(random) $± 0.18 $(systematic) mag]{shappeestanek11}), and reddenings $E(B-V) =
0.014$\,mag in the host galaxy and $E(B-V) = 0.009$\,mag in the Milky Way
\citep{schlegel98,Patat13}. 

\begin{figure*}
 \includegraphics[width=178mm]{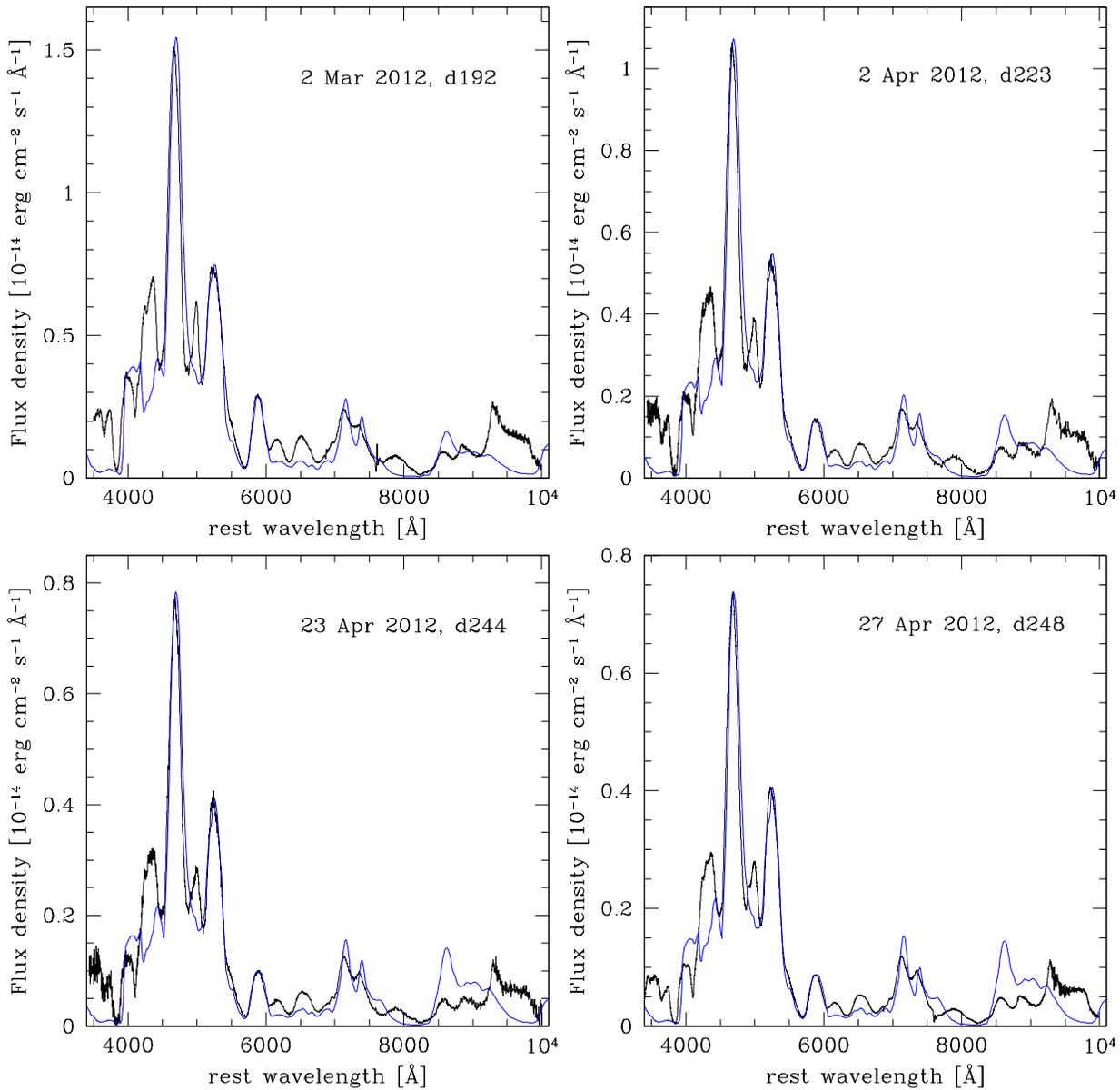}
\caption{Synthetic spectra based on the $\rho$-11fe model (blue) 
compared with observed spectra (black): earlier spectra.}
\label{fig:allspec_1}
\end{figure*}

\begin{figure*}
 \includegraphics[width=178mm]{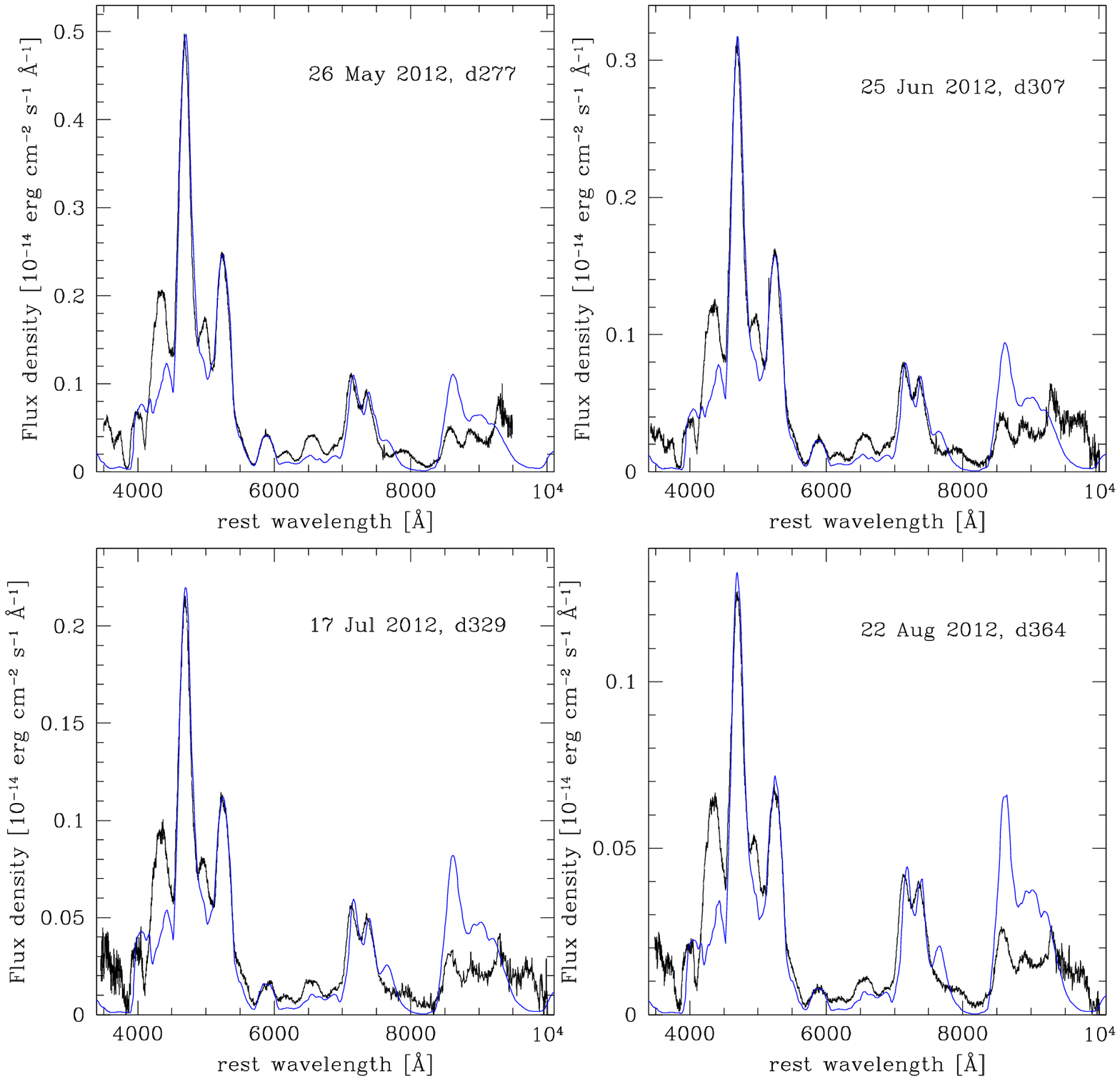}
\caption{Synthetic spectra based on the $\rho$-11fe model (blue) 
compared with observed spectra (black): later spectra.}
\label{fig:allspec_2}
\end{figure*}

We find that the nebular spectra of SN\,2011fe require on average a \Nifs\ mass
of $\sim 0.47\, \Msun$. This is consistent with previous estimates based on the
peak of the light curve \citep{pereira13}. Our set of nebular models is shown
in Figures \ref{fig:allspec_1} and \ref{fig:allspec_2}, and Table 1 summarises
the main results. There seems to be a tendency for the \Nifs\ mass to increase
for later-time spectra.  This may be caused by inaccuracies in our description
of positron escape, since little is known about positron deposition
\citep{cappellaro97}, but in test models with full positron trapping we are
getting similar results. Another possibility is that too much IR flux is
produced by our models for the later epochs. There may be indications of this
in the Fe-dominated emission features in the red, which are increasingly
overestimated by the model spectra, while the emission at 9400--9800\,\AA\ is
not reproduced in our models. These are high-lying \FeII\ transitions. One
possibility is that the collision strengths of these and other weak lines are
not well determined. Overall, however, given the poor fitting of some features,
possible uncertainties in flux calibration, and the neglect of
radiative-transfer effects, we regard our \Nifs\ mass estimate to be within the
uncertainties.  

Another region that is poorly reproduced is the emission near 4250\,AA. Several
moderately strong [\FeII] lines are present there, including lines at 4244\,AA, 
4277\,AA, and 4416\,AA, but their combined strength is only about half of what
is observed. This may be due to problems with collision strengths, A-values, or
it may indicate the presence of some other species which is not considered in
our models. The same is true for the narrow emission near 5000\,AA.

The element distribution determined from the early-time spectra is compatible
with the late-time models. These are really only sensitive out to $v \approx
8000$\,\kms, which is the width of the emission lines. Fitting the Fe emission
lines, we find a roughly constant width, perhaps with a slight tendency to
decrease with time. The characteristic width at epochs of $\sim 1$\,yr is $\sim
7600$\,\kms. Based on a relation between emission-line width and light-curve
shape \citep{mazzali98}, we expect for SN\,2011fe a full width at half-maximum
intensity of the strongest Fe line blend of $\sim$\,12,500\,\kms, and this is
indeed what we measure\footnote{The width of the emission line reflects both
velocity broadening and the blending of several lines that contribute to the
feature \citep{mazzali98}.}. This corresponds to a normal SN\,Ia, but appears
to be somewhat narrow for the \Nifs\ mass and the light-curve shape 
\citep[$\Dm = 1.07 \pm 0.06$\,mag;][]{mcclelland13}.

\begin{figure*}
 \includegraphics[width=133mm]{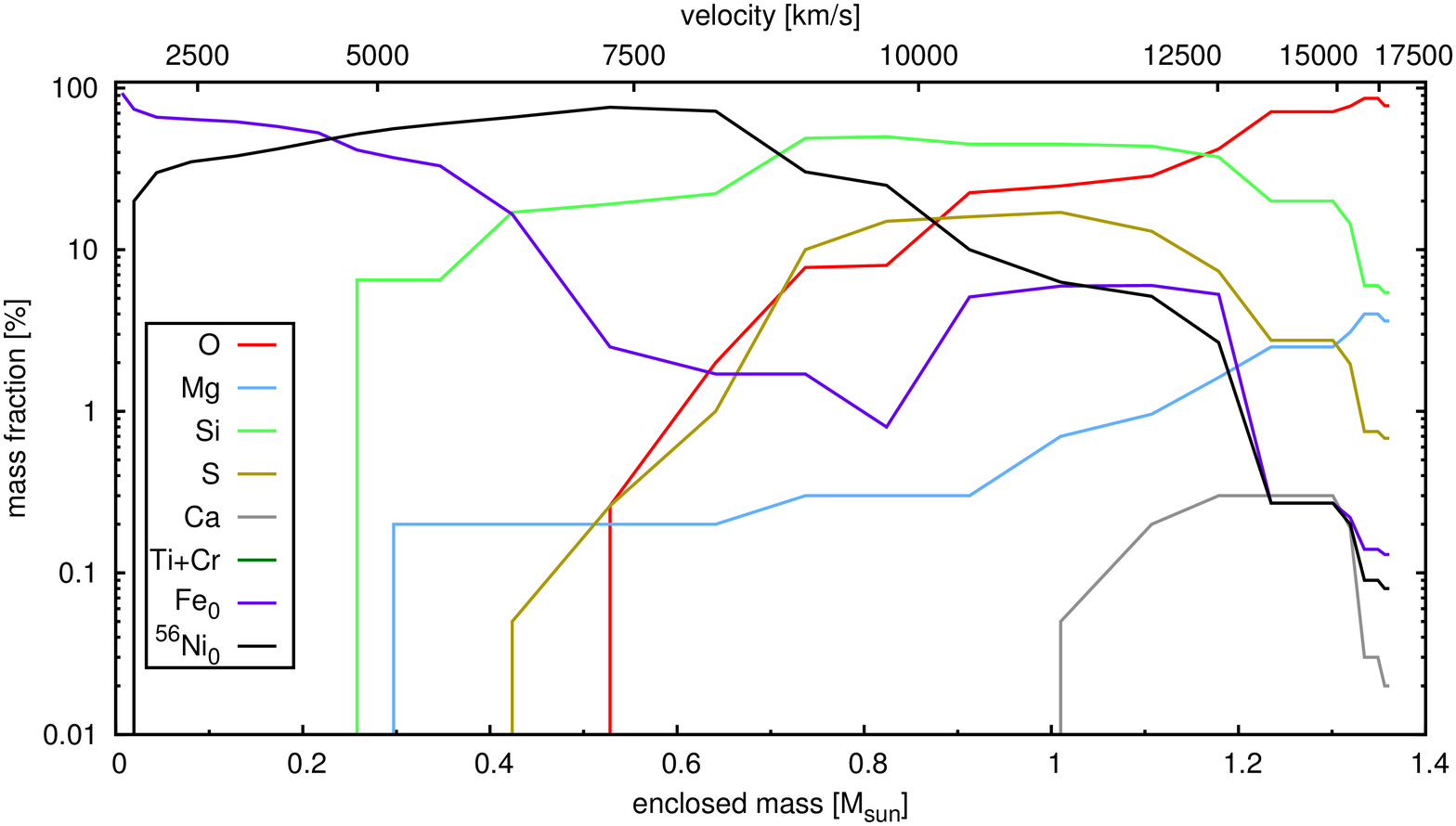}
 \includegraphics[width=133mm]{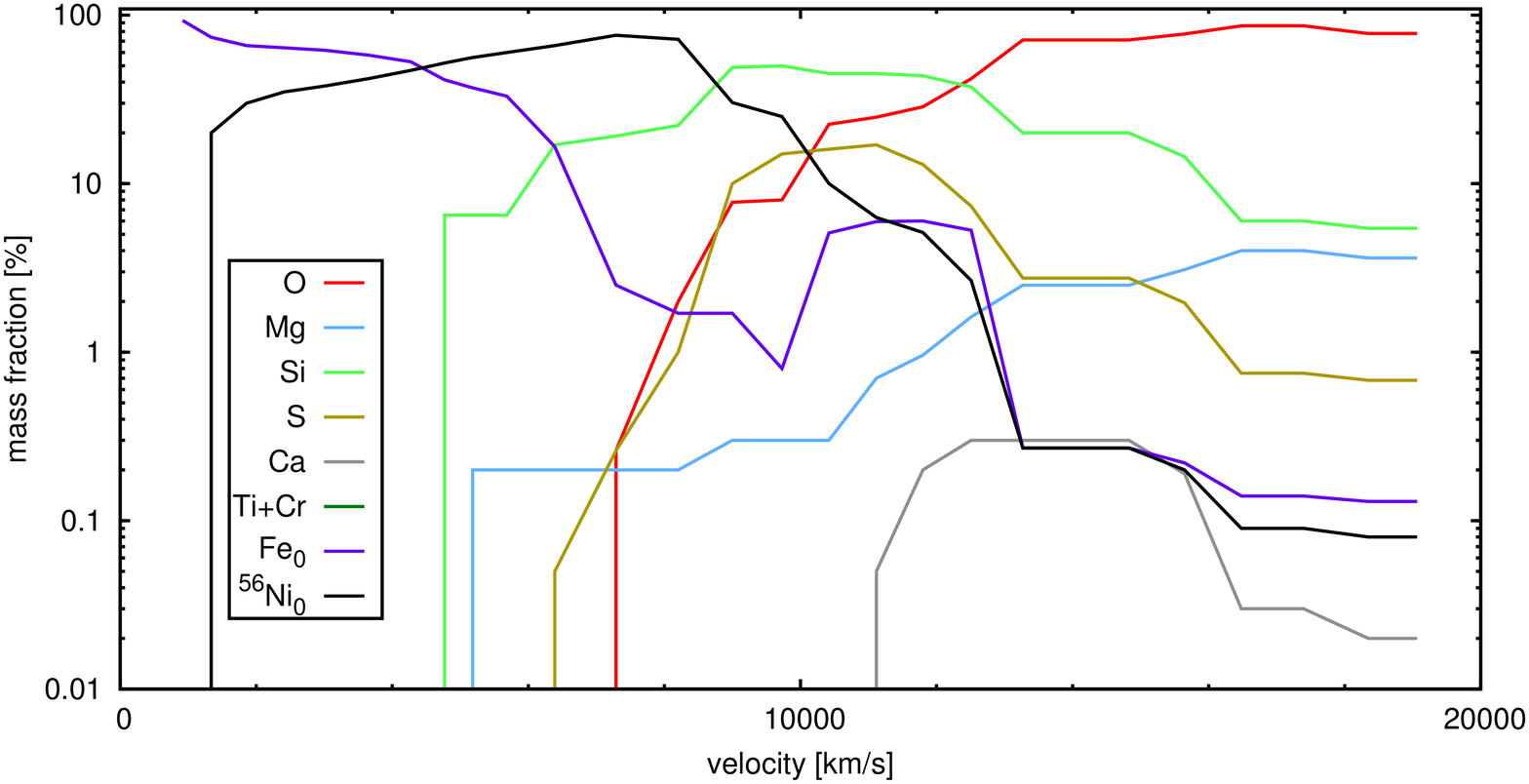}
\caption{Abundance distribution in the ejecta of SN\,2011fe. The elements that
are shown are those that can be inferred from spectroscopic modelling of
\citet{mazzali14}. 
The top panel is linear with respect to mass, the bottom panel with respect 
to velocity.}
\label{fig:Abundances}
\end{figure*}

Our total \Nifs\ mass is only slightly larger than the mass determined from our
early-time analysis at velocities $v > 4500$\,\kms. Only an additional $\sim
0.1\, \Msun$ of \Nifs\ is found to be located at $v < 4500$\,\kms. As we
mentioned, model $\rho$-11fe has $\sim 0.25\, \Msun$ of material below
4500\,\kms. This means that $\sim 0.15\, \Msun$ of material at low velocity is
not \Nifs, but rather stable nuclear statistical equilibrium (NSE) species. Two
major species are predicted to be synthesised in n-rich burning: \Feff\ and
\Nife, both of which are stable isotopes. We have a direct handle on the
abundance of stable Ni from the emission line near 7400\,\AA, which is normally
identified as [\NiII] $\lambda\lambda$7380, 7410. Indeed, the small emission
seen in SN\,2011fe near 7400\,\AA\ can only be reproduced if some Ni is present
at the lowest velocities. Given the advanced epoch, this must be stable Ni,
presumably \Nife. This is produced from \Feff\ in the so-called alpha-rich
freezeout phase via \Feff\ $(\alpha, \gamma$) \Nife\ during the early
expansion. However, the broad emission in that region is dominated by other
lines, [\FeII] $\lambda\lambda$7388, 7452 and [\CaII] $\lambda\lambda$7291,
7324. Only a very small mass of \Nife\ is required to reproduce the small
emission that corresponds in wavelength to the [\NiII] lines. On average, a
value of $\sim 0.008\, \Msun$ of \Nife, confined below $v \approx 4000$\,\kms\
and concentrated at the lowest velocities, is sufficient. Even at the lowest
velocities, the abundance of \Nife\ never exceeds 10\%.  This means that most
of the mass at low velocities is composed of Fe. Again, this Fe cannot all come
from \Nifs\ decay, because if it did the SN luminosity would be too high. We
presume that this is mostly \Feff. 

Although we cannot distinguish \Feff\ from directly synthesised \Fefs, the
latter is supposed to be produced at low $Y_e$, inside the \Nifs\ region
\citep{iwamoto99}, which is not where we find it to be located in our
modelling.  This is a common feature in SNe\,Ia \citep{stehle05, mazzali08,
tanaka11, sasdelli14}, and it may indicate a higher $Y_e$ than predicted by
models such as $W7$. The presence of stable Fe helps keeping the ionisation
degree sufficiently low that the correct relative contribution of [\FeII] and
[\FeIII] lines is achieved in our models. We showed in other work that if this
central density is low, ionisation becomes too high and [\FeIII] lines
dominate, leading to a strong 4800\,\AA\ emission \citep{mazzali11}. If the
density is even lower, only [\FeIII] lines are produced and the emission-line
spectrum changes nature. This seems to be a feature of the fastest-declining
SNe\,Ia \citep{mazzali12}. The total mass of stable Fe (obtained by adding
material already diagnosed at high velocities by \citealt{mazzali14}, which is
necessary to reproduce Fe lines at early times, when \Nifs\ has not yet decayed
to \Fefs, and stable Fe at low velocities as determined here) is on average
$\sim 0.23\, \Msun$. This is in line with the results of \citet{mazzali07a} and
\citet{kasenwoosley07}. 

A plot of our abundance distribution is shown in Figure \ref{fig:Abundances}.
This compares favourably with one-dimensional (1D) Chandrasekhar-mass models
based on the SD scenario, where ignition occurs near the centre of the WD at
high density, and a sufficiently high temperature is achieved that n-rich
isotopes can be synthesised.

\subsection{The Near-Infrared}

\begin{figure*}
 \includegraphics[width=134mm,angle=-90]{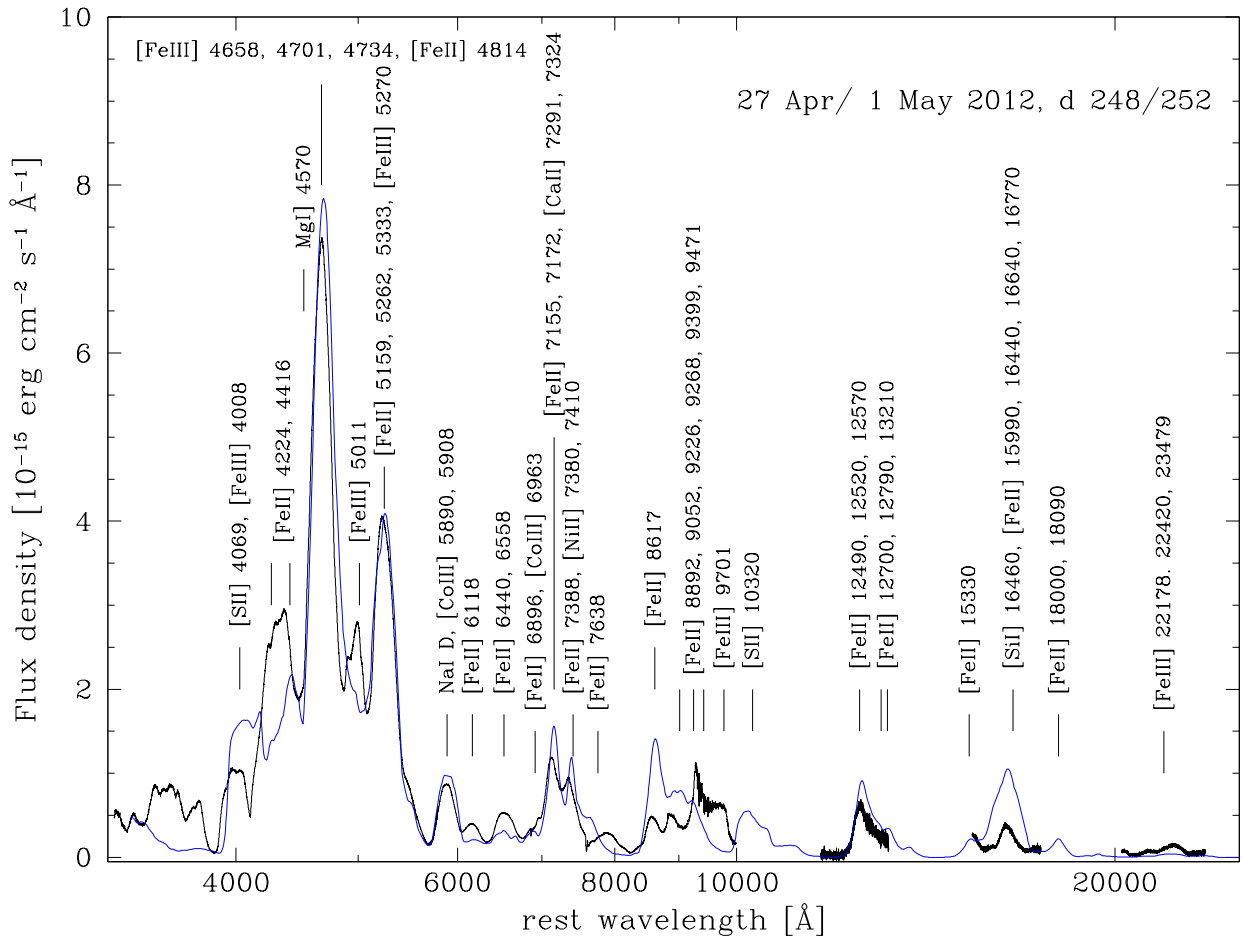}
\caption{Synthetic spectrum based on the $\rho$-11fe model (blue), now also showing 
 the IR. The observed IR spectrum (1 May 2012) is given along with the optical 
part (27 Apr 2011). The strongest lines in the model, as well as lines that 
match observed features in wavelength but possibly not strength, are labelled.}
\label{fig:optIR}
\end{figure*}

We mentioned that as time goes on, the flux is expected to shift to the
infrared. We do have one epoch of NIR spectroscopy, obtained with the LBT on 1
May 2012. This spectrum can be combined with the optical spectrum obtained on
27 April 2012. In Figure \ref{fig:optIR} we show our model for that epoch, now
extending to the NIR. The model reproduces the $J$- and $K$-band spectrum
reasonably well, while the flux in the $H$ band is too low. This is in part
because [\SiI] 1.6\,$\mu$m is overestimated, the result of \Nifs\ being in
close contact with Si, but we also seem to somewhat overpredict the Fe emission
in that region. 

It is possible that a sharper velocity separation between \Nifs\ and IMEs may
be required than the results of \citet{mazzali14} suggest. We tested this in a
model, but the resulting spectrum still showed too much flux at 1.6\,$\mu$m.
Using a three-dimensional (3D) approach may be useful, concentrating \Nifs\ in
blobs or clumps from which $\gamma$-rays escape less successfully. On the other
hand, while this may help,  $\gamma$-rays escape quite easily at the low
densities that in general characterise the nebular regime, so the overdensity
would probably have to be quite large. 

The other possibility is that the density in the inner region may still be
somewhat too high. In the case of SN\,2003hv, a sharp reduction of the NIR flux
was obtained when the central densities were decreased, shifting the flux from
[\FeII] to [\FeIII] lines. The ratio of these two ions in the optical is
correctly reproduced for SN\,2011fe, however, so we do not expect that the
density can be reduced by a large amount. In any case, we can test this
possibility using sub-Chandrasekhar-mass models.

\subsection{sub-Chandrasekhar-mass models}

\begin{figure*}
 \includegraphics[width=134mm,angle=-90]{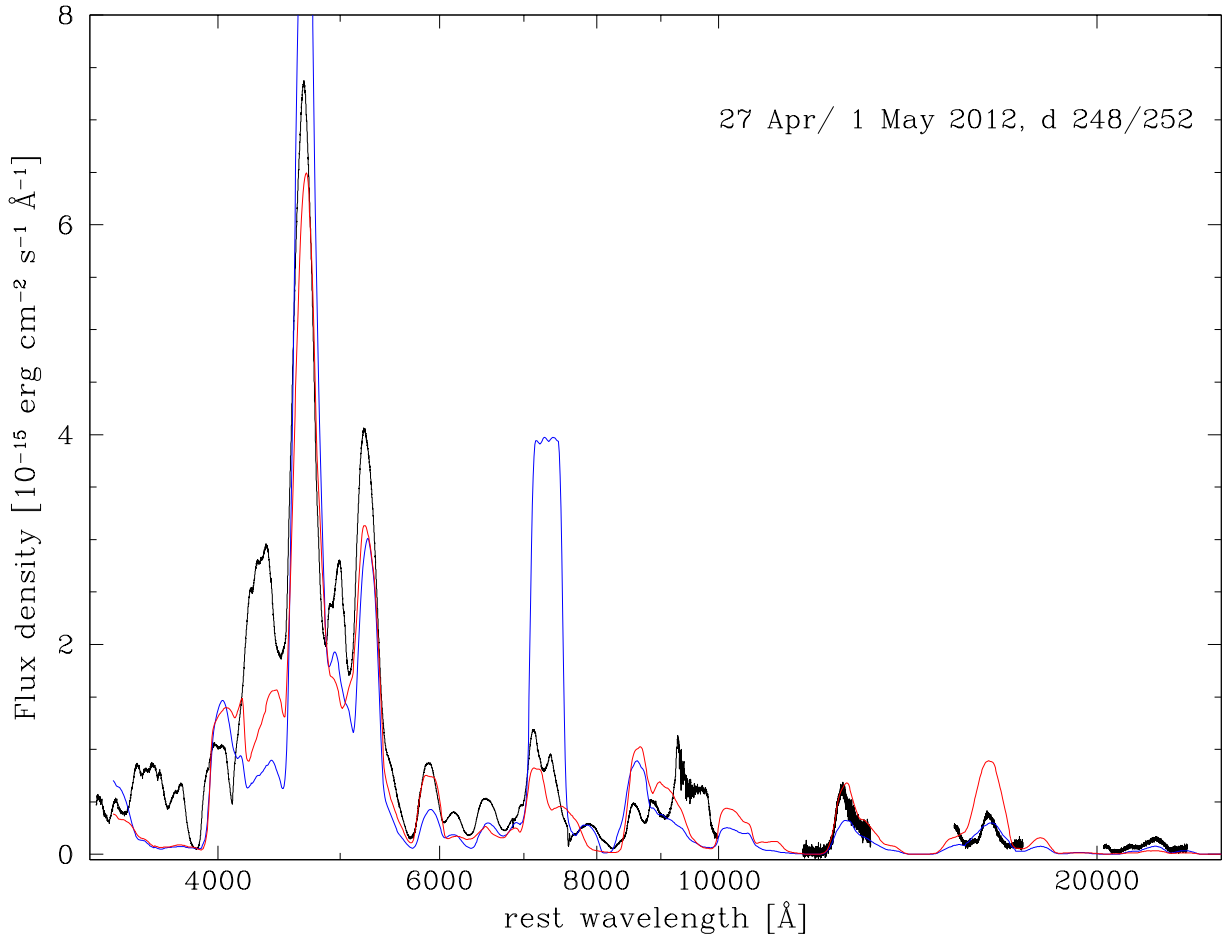}
\caption{Sub-Chandra models: CDT3 (blue) and a model with $M = 1.26\,\Msun$ (red). 
The observed spectrum is shown in black.}
\label{fig:subcha}
\end{figure*}

In order to verify the need for a core of stable NSE elements, we tested a
sub-Chandrasekhar-mass model. In such models, explosive burning begins at the
outside, typically following the accretion of He-rich material from a
companion. Most models suggest that a secondary detonation can be triggered
within the WD, which can be incinerated even if its mass is significantly below
the Chandrasekhar limit \citep{livnearnett95,fink10}. However, because the
density never reaches the high values typical of near-Chandrasekhar-mass WDs,
n-rich nucleosynthesis is not expected to take place. Models are characterised
by low central densities. 

We used the density and abundance distribution of a 1D sub-Chandrasekhar-mass
model from the set presented by  \citet{shigeyama92}. We selected model CDT3,
which yields a \Nifs\ mass of $0.50\, \Msun$, in agreement with the estimated
value. The total mass of the model is $\sim 1\, \Msun$. The mass of stable
Fe-group isotopes, \Feff\ and \Nife, is very small, $\sim 0.003\, \Msun$ each.
Intermediate-mass elements are present with significant abundances: Si, $\sim
0.21\, \Msun$ and S, $\sim 0.12\, \Msun$. Some unburned oxygen is also present,
$\sim 0.07\, \Msun$. The density structure of this model is shown in Figure
\ref{fig:densities}. 

The corresponding synthetic spectrum computed for day 248 is shown in Figure
\ref{fig:subcha}. It is characterised by high ionisation, as shown by the ratio
of [\FeIII] and [\FeII] lines: the [\FeIII]-dominated emission near 4800\,\AA\
is much stronger than the [\FeII]-dominated emission near 5200\,\AA.  This is
reminiscent of SN\,2003hv, and is the result of the low central density
\citep{leloudas09,mazzali11}.  Elsewhere, the small Si mass, its separation
from \Nifs, and the lack of \FeII\ means that the synthetic line at $\sim 1.6\,
\mu$m is now weaker than the observed one. However, other synthetic [\FeII]
lines, such as those near 8800\,\AA, are still too strong, suggesting a problem
with collision strengths or, less likely, $A$-values. On the other hand, the
[\NiII] line is no longer reproduced. Also, very noticeably, the Ca abundance
is much too high, as shown by the strength of the emission near 7200\,\AA. This
demonstrates that a model which is significantly sub-Chandrasekhar is not
appropriate for SN\,2011fe. It is much more difficult to establish whether a
model which is just below the Chandrasekhar mass can also be ruled out. This
mostly depends on the exact nucleosynthesis. It also confirms that stable Fe is
required in order to reproduce the spectra. 

As an alternative, we constructed a sub-Chandrasekhar model with a larger mass.
In this test model the total mass is $1.26\, \Msun$, the stable Fe content is 
small but not zero ($0.12\, \Msun$), and the \Nife\ mass is $0.43\, \Msun$,
consistent with our results above. The synthetic spectrum (Figure
\ref{fig:subcha}) shows again the problem of high ionisation, although to a
lesser extent. The best results is obtained when the mass approaches \MCh\ and
the stable Fe content is close to $0.2\, \Msun$.

\subsection{Where is the stable Fe?}

\begin{figure*}
\includegraphics[width=134mm,angle=-90]{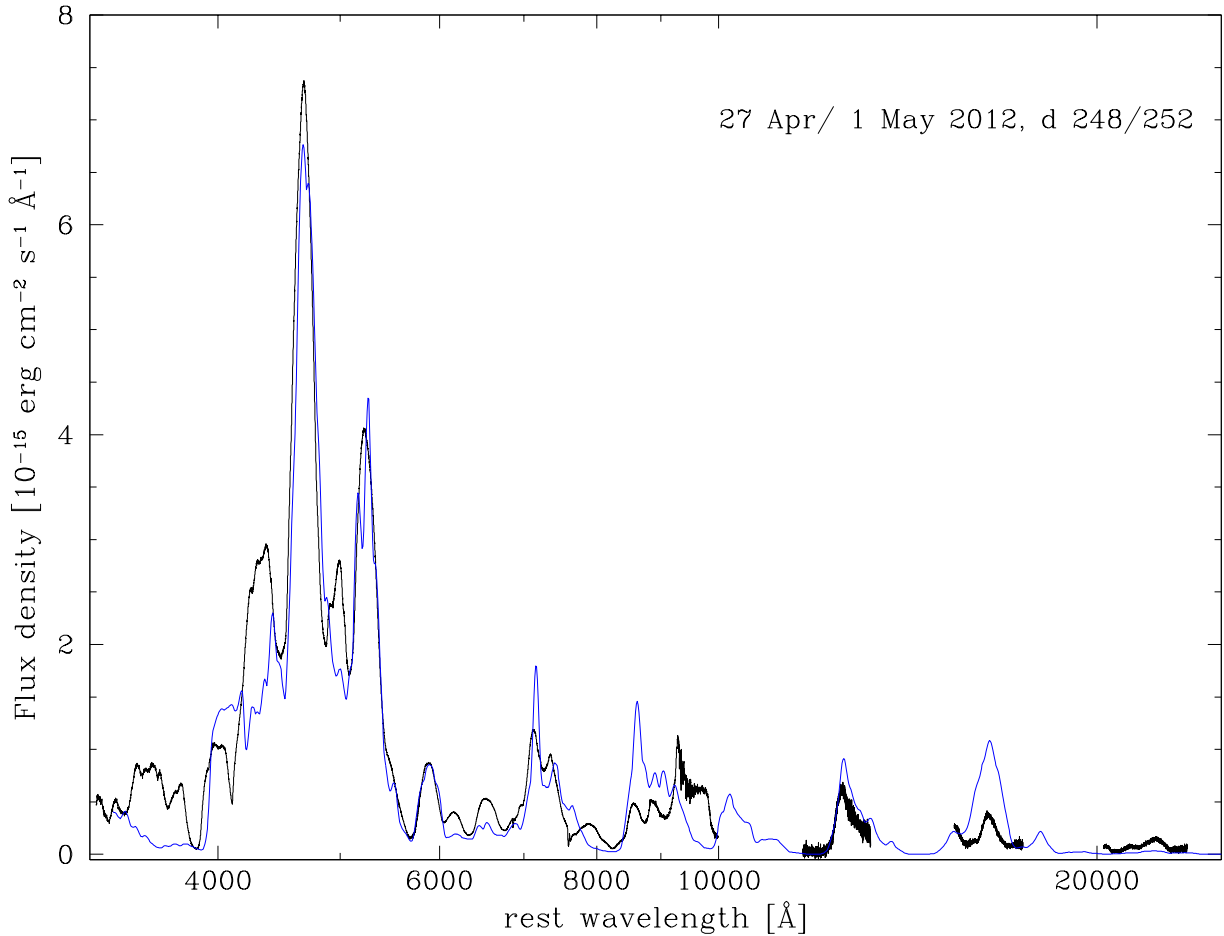}
\caption{Synthetic spectrum for the Ni-Fe inversion model (blue). The observed 
spectrum is shown in black.}
\label{fig:NiFeinvert}
\end{figure*}

Having established that a high central density is needed, and that the presence
of a significant amount of \Feff\ and stable \Fefs\ produced at burning
\citep{iwamoto99} appears to be necessary, we can address the question of its
location. Recent 3D delayed-detonation models suggest that  n-rich NSE species
produced in the deflagration phase rise because of their buoyancy and end up
expanding at higher velocities than \Nifs\ when the simulation is performed in
3D. This leads to an inverse structure, where \Nifs\ is located inside n-rich
NSE species \citep{seit13}\footnote{This did not seem to have been noticed in
earlier 3D simulations \citep{gamezo05}.}. In order to assess the effect of
this, we simulated spectrum formation using the $\rho$-11fe density structure
but moving stable Fe and Ni above \Nifs\ in velocity. This process requires a
moderate rescaling of the \Nifs\ mass to obtain the same luminosity, because
when \Nifs\ is located deeper in the ejecta $\gamma$-rays deposit more
efficiently. The flux level of the nebular spectrum can be reproduced for a
\Nifs\ mass of only $0.34\, \Msun$. The mass of stable Fe  becomes $0.43\,
\Msun$, and \Nife\ slightly increases to $0.01\, \Msun$. The resulting model is
shown in Figure \ref{fig:NiFeinvert}.

Although the model has reasonable features, three areas cause concern. First,
all Fe lines are narrow.  This is because there is too little \Nifs\ at high
velocity, and the Fe that is located there is not efficiently heated by the 
$\gamma$-rays, which are emitted very deep in the ejecta. Increasing the mass
of \Nifs\ would help, but the flux would be too high. Additional mixing may fix
this problem, but it may take us back toward a situation where \Feff\ is below
\Nifs. Second, the ratio of Fe lines now favours \FeII.  This is not
surprising, since radiation comes predominantly from the densest, innermost
regions, where recombination is most effective. The ratio of the two strongest
emission lines, whose atomic data are probably best known, is a driving
parameter to establish the success of a model. Finally, the \Nifs\ mass is in
rather strong disagreement with the value derived from the peak of the light
curve. Combining the smaller \Nifs\ mass with the fact that the diffusion time
of the optical photons can only increase if they are produced deeper in the
ejecta, as would be the case if \Nifs\ were located at the lowest velocities,
this ad-hoc model is not expected to reproduce the observed light curve. 

In conclusion, the nebular spectrum is not as well reproduced by this model as
it is when \Nifs\ is located outside stable Fe. Also, whether such a small
\Nifs\ mass is able to reproduce the peak of the light curve is highly
questionable. We tend not to favour this mock model. Tests should be performed
with the real model, which is obviously technically superior to the 1D
calculation.

\section{Light-curve model}

The description of the innermost part of the ejecta deduced through models of
the late-time spectra can be combined with that of the outer layers obtained
through the analysis of the early-time spectra. Figure \ref{fig:Abundances}
gives the abundance distribution, while the density profile was shown in Figure
\ref{fig:densities}. Although the nebular results seem to confirm the
distribution in density that was derived from the early-time data, the early-
and late-time analyses are somewhat disconnected. A test for the overall model,
in the spirit of abundance tomography
\citep{stehle05,mazzali08,tanaka11,sasdelli14}, can be provided by the
independent study of the light curve. The density and abundance stratification
can be used to compute the emission and deposition of $\gamma$-rays and
positrons from \Nifs\ and \Cofs\ decay, the propagation of optical photons, and
the final emission of radiation. We employ a Monte Carlo code developed by
P.A.M. and first used by \citet{cappellaro97}. The opacity description follows
\citet{mazzali01}, with different weights assigned to different element groups
\citep[see][]{khokhlov93} in order to describe the dominance of line opacity in
a SN\,Ia \citep{pauldrach96}. 

\begin{figure}
 \includegraphics[width=88mm]{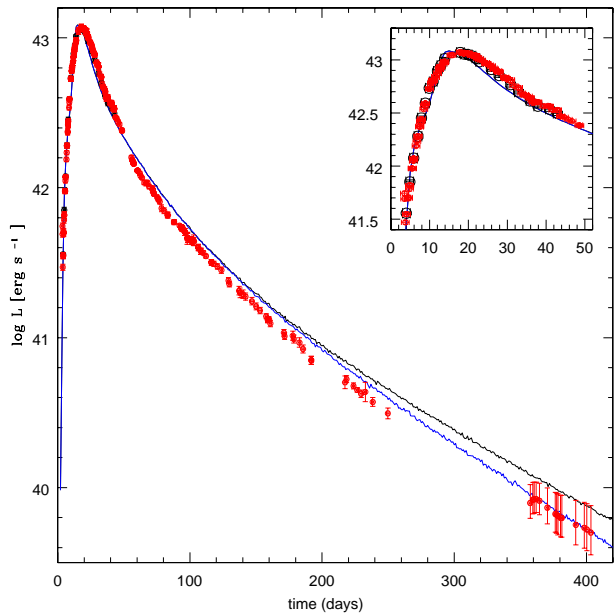}
 \caption{Synthetic bolometric light curves based on the density of $\rho$-11fe
and the abundance stratification derived here and by \citet{mazzali14}. The 
UV through IR pseudo-bolometric light curve computed by \citet{pereira13} is shown
as black circles and ours as red circles. The uncertainties in the flux are as
quoted in that paper. An equivalent bolometric light curve computed by S.B. is
shown as red circles. The epoch of both bolometric light curves has been shifted
to match a risetime of 18.3 days, following \citet{mazzali14}. The lines show
two light curves computed with our density and abundance results: 
$\tau_{e^+} = 7$\,g cm$^{-2}$ (black) and
$\tau_{e^+} = 2$\,g cm$^{-2}$ (blue).}
\label{fig:bolLC}
\end{figure}

The resulting bolometric light curve is shown in Figure \ref{fig:bolLC}. There
we also give a pseudo-bolometric light curve computed based on the available UV
through IR photometry. It is very similar to the light curve published by
\citet{pereira13} in the region of overlap but it extends to later epochs. The
epochs of the observed light curve were offset to match the explosion date
derived by \citet{mazzali14} --- 22 Aug. 2011, MJD55795.3 --- implying a rise
time of 18.3 days. The agreement between the synthetic light curve and the
observed one is quite good, both in flux level (which confirms the \Nifs\ mass,
the density, and the abundance stratification) and in its development over time
(which is a result of photon diffusion and again tests density, \Nifs\ mass,
and abundance distribution). The agreement in the rising part of the light
curve is remarkable. This would have been worse had we adopted a shorter rise
time. More mixing out of \Nifs\ might have been required in that case, but this
would be in conflict with results from the early-time spectra
\citep{mazzali14}. The main shortcomings of our synthetic light curve are the
slight underestimate of the flux just after maximum brightness (see Figure
\ref{fig:bolLC}, inset) and the overestimate at 100--200 days. The discrepancy
just after peak may indicate that we are underestimating the opacity in the
inner ejecta, which is quite possible given our rather coarse grey treatment.
The discrepancy at day 100--200 may be due to the opposite problem, but also to
a less well determined correction for flux outside the optical bands, which may
shift the bolometric light curve of SN\,2011fe upward.


\section{Discussion}

Our results appear to confirm that the density structure derived by
\citet{mazzali14} can appropriately describe SN\,2011fe when it is extended to
low velocities under the assumption that the explosion ejected approximately
one Chandrasekhar mass of material. They also strongly favour the presence of a
significant amount of \Feff, as long as it is located at the lowest velocities.
The mass of \Nifs\ produced by SN\,2011fe is $\sim 0.45\, \Msun$, in agreement
with estimates from the light curve, if \Nifs\ is located above \Feff. On the
other hand, if \Nifs\ is located centrally, the mass which is necessary to
power the late-time spectra is significantly smaller. The amount of stable
\Nife\ required to match the observed emission near 7400\,\AA\ is very small,
$\sim 0.06\, \Msun$, because that emission feature is dominated by [\FeII] and
[\CaII] lines. 

While it is difficult to determine the mass with high precision (even in the 
case of SN\,2011fe the uncertainties in luminosity, combined with the intrinsic 
approximations of the model, mean that we cannot claim better than 20\% 
accuracy), we note that the kinetic energy of model $\rho$-11fe matches
very well the expected \KE\ yield based on the nucleosynthesis. Applying
an equation relating the yield of different species to the \KE\ produced in a
Chandrasekhar-mass SN\,Ia explosion,
\begin{eqnarray}
E_{\rm K} = [1.56\, M(^{56}{\rm Ni}) + 1.74\, M({\rm stable NSE}) + \nonumber \\ 
	  1.24\, M({\rm IME}) - 0.46] \times 10^{51}\,{\rm erg},
\end{eqnarray}
\citep{woosley07}, where the constant term is the binding energy of the WD, 
and inserting $M$(\Nifs) = 0.47\,\Msun, $M$(stable NSE) = 0.24\,\Msun, and
$M$(IME) = 0.41\,\Msun, we obtain \KE $= 1.20 \times 10^{52}$\,erg, in excellent 
agreement with the energy of $\rho$-11fe, $1.32 \times 10^{52}$\,erg. This 
suggests that a weak delayed detonation of a near-Chandrasekhar-mass white 
dwarf is a consistent model for SN\,2011fe. The fact that the emission-line
width of SN\,2011fe is low for its light-curve shape (the width of the
strongest lines is very similar to those of SNe\,1992A and 2004eo, which had
$\Dm \approx 1.4$\,mag) also indicates that the specific energy of SN\,2011fe 
was somewhat lower than normal.

Other points are worthy of notice, as follows.

(1) 1D--3D. Our 1D model $\rho$-11fe resembles delayed-detonation models that
were computed in 1D, such as those of \citet{iwamoto99}.  1D delayed-detonation
models predict that n-rich NSE species are located deeper than \Nifs. Recently,
a 3D delayed-detonation model has been published which suggests that \Nifs\ is
located below stable Fe \citep{seit13}. A test with a simplistic reproduction
of that situation does not lead to an equally good synthetic spectrum. There
seems to be no reason why our 1D synthetic spectra should be biased in favour
of 1D explosion models, as these are computed completely independently. 


(2) Mass of progenitor. An accurate estimate of the mass does not seem to be
possible to better than $\sim 20$\%, because of uncertainties related to data
calibration, distance and reddening, as well as to the modelling process.
However, our models require the presence of n-rich NSE species at the deepest
velocities in order to achieve an ionisation regime that allows the ratio of
the strongest [\FeIII] and [\FeII] lines to be reproduced. This is natural in
SD models where the progenitor WD is approaching the Chandrasekhar mass and has
a high central density. It is not in edge-lit sub-Chandrasekhar-mass models, or
in merger models, where in both cases the central density is lower. These two
classes of events may give rise to rather distinct spectroscopic signatures in
the nebular phase, which could explain some subtypes of SNe\,Ia
\citep{mazzali11,mazzali12}. On the other hand, we cannot exclude a mass
somewhat different from the Chandrasekhar mass, or a different density profile.
In particular, the nebular lines are narrow for a SN with the light-curve shape
of SN\,2011fe ($\Dm \approx 1.1$\,mag).  The model we have presented here seems
to show a high degree of consistency.  It needs to be tested whether
first-principle 3D models can explain SN\,2011fe. This could be done using
codes such as those of \citet{maeda06}. 

(3) The Ni line. If the line near 7400\,\AA\ is indeed [\NiII]
$\lambda\lambda$7380, 7410, a small amount of \Nife\ is sufficient to reproduce
it. A possible blueshift \citep[$\sim 1100$\,\kms;][]{mcclelland13} is observed
with respect to the rest-frame wavelength of the lines, after correction for
cosmological redshift. In the aspherical explosion scenario of \citet{maeda10}, 
this is consistent with SN\,2011fe being a low velocity gradient (LVG) SN 
\citep[$\dot{v} = 35 \pm 5$\,\kms d$^{-1}$;][]{mcclelland13}. 
It should be noted that the [\NiII] line is blended with the emission near 
7300\,\AA, which is dominated by [\CaII] and [\FeII] lines. Some of the 
blueshift of the Ni line may simply be the result of this blending.

One of the ingredients of the 3D model sketched by \citet{maeda10} is that the
centrally located n-rich NSE species are not excited by $\gamma$-rays, and that
this would be revealed by flat-topped profiles of some emission lines. In
particular, in the case of SN\,2003hv, evidence of flat tops was found in the
NIR spectrum \citep{motohara06,gerardy07}.  However, our models show that it is
not difficult for inwardly propagating $\gamma$-rays emitted from a \Nifs\ zone
that surrounds a nonradioactive central core to excite material located at the
lowest velocities. This is indicated by the sharp peak of the [\NiII] emission.
Also, in the case of SN\,2011fe, the NIR spectrum does not seem to show any
flat-topped profiles. This suggests that not only is \Nife\ irradiated, but
\Feff\ is as well. In particular, the line at $1.26\, \mu$m is dominated by a
single line, [\FeII] $1.257\, \mu$m, and hence is the perfect candidate for a
flat top, but we see no evidence for one. Incidentally, a flat top in that line
does not seem to be visible in the spectrum of SN\,2003hv either
\citep{motohara06}, and the emission could be reproduced quite well with a 1D
model \citep{mazzali11}. The other line for which a flat top was observed in
SN\,2003hv is at $\sim 1.6\, \mu$m. This line is a blend of [\FeII] and
[\SiII]. It is not impossible that the observed flat top is the result of the
blending of these lines and the relatively low signal-to-noise ratio of the NIR
spectrum of SN\,2003hv. In SN\,2011fe no flat top is present in this feature,
but SN\,2011fe was more luminous than SN\,2003hv, and therefore it probably
produced less Si and other IMEs \citep{mazzali07a}. So, if the flat tops in
SN\,2003hv are real, the density contrast between the n-rich core and the
surrounding \Nifs\ shell was perhaps very large, so that $\gamma$-rays could
not penetrate. This may be unlikely if SN\,2003hv was a sub-Chandrasekhar-mass
explosion \citep{mazzali12}, but it may be worth testing. 

(4) Hydrogen? No narrow H$\alpha$ emission is seen in the nebular spectra
\citep[see also][]{shappee13}. An emission line is observed at a similar
wavelength, and the model indicates that it may be [\FeII] $\lambda$6559, but
the synthetic line does not match the observed one in flux. Several other weak
lines are poorly reproduced. They are mostly attributed to Fe lines because of
wavelength coincidence (see Fig. \ref{fig:optIR}). Their incorrect strengths
may be caused by poorly known collision strengths. However, if the distribution
of H stripped from a companion was not as narrow as predicted by some models
\citep{marietta00,pakmor11}, H$\alpha$ could make a contribution to the
observed emission. The observed line has similar width as all other lines.

\section{Conclusions}

We have presented 1D models for a series of optical and one NIR spectrum of the
nearby, spectroscopically normal SN\,Ia 2011fe. We have shown that if we use
the density distribution derived from the early-time optical/UV spectra
\citep[``$\rho$-11fe";][]{mazzali14}, and extend it to low velocities assuming
that the ejected mass equals the Chandrasekhar mass, we can obtain a
satisfactory match of the main features of the late-time spectra using a \Nifs\
mass of $\sim 0.45\, \Msun$ and a stable Fe mass of $\sim 0.23\, \Msun$. Stable
Fe should be mostly located in the central, lowest-velocity region of the
ejecta for a best match to be obtained. This is consistent with 1D 
delayed-detonation models \citep[\eg][]{iwamoto99}, but not with a recent 3D
model \citep[but see Gamezo et al. 2005]{seit13}. The derived density and
abundance distributions were used to compute a bolometric light curve, in the
spirit of abundance tomography. The result compares favourably with the
pseudo-bolometric light curve of SN\,2011fe \citep{pereira13}, lending further
credence to the model proposed here.

A significantly sub-Chandrasekhar model seems to be ruled out because the low
central density and the ensuing lack of newly synthesised \Feff\ lead to a high
ionisation and incorrect ratios of the strongest [\FeII] and [\FeIII] lines.
Merger models that involve WDs significantly below the Chandrasekhar mass may
have similar characteristics, and would then be equally disfavoured on these
grounds. 

A weak line near 7400\,\AA\ can be reproduced as [\NiII], requiring the presence
of a small amount of \Nifs\  ($\sim 0.06\, \Msun$). The line shows a blueshift
of $\sim 1100$\,\kms\ with respect to its rest wavelength, in agreement with the
LVG nature of SN\,2011fe. The central Fe region is effectively excited by
$\gamma$-rays from the \Nifs\ region that surrounds it, preventing the formation
of flat-top emission profiles even for rather isolated Fe lines. 


The evidence concerning the nature of SN\,2011fe remains ambiguous
\citep{roepke12}. Spectral fits seem to favour a model resembling a 1D delayed
detonation, but with lower energy \citep[this paper;][]{mazzali14}. On the
other hand, much observational evidence, including the lack of hydrogen in the
late-time spectra \citep{shappee13}, the lack of X-ray or radio detections
\citep{horesh12}, the absence of early-time extra flux in the light curve
\citep{bloom12}, and the lack of an imaged companion \citep{li11}, does not
support SD models with a main sequence or red giant companion. On the other
hand, models that do not reach high central densities and do not produce stable
NSE isotopes, including significantly sub-Chandrasekhar models (such as might
result from the accretion of helium from a nondegenerate He companion with
small radius) or mergers of WDs whose individual mass is well below the
Chandrasekhar mass, are not favoured by late-time models. Additionally, no
evidence for any residual He was seen in the early-time spectra
\citep{parrent12,mazzali14}. On the other hand, a marginally sub-Chandrasekhar
model may be able to reproduce both the low energy and the presence of stable
NSE isotopes. Alternatively, a marginally super-Chandrasekhar model may also
display all observed features. 


In summary, spectroscopic evidence seems to favour for SN\,2011fe a model with
a high central density, matching the structure of a low-energy delayed
detonation  of a Chandrasekhar-mass WD in a SD system. The model we used has
$\KE = 1.20 \times 10^{51}$\,erg. Currently, 1D models offer a better match to
the data than 3D models, but this should be taken with some caution because 1D
models are intrinsically simpler. Further work on this and other models would
be useful to bring them into agreement with all observed properties of
SNe\,Ia.

\section*{ACKNOWLEDGEMENTS}

We would like to thank S. Bradley Cenko, Isaac Shivvers, Ori Fox, Pat Kelly, 
and Daniel Cohen for assistance with the observations. 
M.S. acknowledges support from the Royal Society and EU/FP7-ERC grant number
[615929]. 
A.V.F. is grateful for support from 
the Richard \& Rhoda Goldman Fund, the Christopher R. Redlich Fund, 
the TABASGO Foundation, and NSF grant AST--1211916.
S.H. is supported by the German Ministry of Education and Research (BMBF) via a
Minerva ARCHES award. 
B.S. is a Hubble, Carnegie-Princeton Fellow, and supported by NASA through
Hubble Fellowship grant HF-51348.001, awarded by the Space Telescope Science
Institute, which is operated by the Association of Universities for Research in
Astronomy, Inc., for NASA, under contract NAS 5-26555. 
J.M.S. is supported by an NSF Astronomy and Astrophysics Postdoctoral Fellowship
under award AST-1302771.
SB is partially supported by the PRIN-INAF 2014 with the project ``Transient 
Universe: unveiling new types of stellar explosions with PESSTO''.
The William Herschel Telescope is operated on the island of La Palma by the
Isaac Newton Group in the Spanish Observatorio del Roque de los Muchachos of 
the Instituto de Astrofísica de Canarias; we are grateful to the staff
there, as well as at Lick Observatory, for their excellent assistance.

\bibliographystyle{mn2e}

\end{document}